# High spatially sensitive quantitative phase imaging assisted with deep neural network for classification of human spermatozoa under stressed condition

Ankit Butola[1, 2#], Daria Popova[2, 3#], Dilip K Prasad[4], Azeem Ahmad[2], Anowarul Habib[2], Jean Claude Tinguely[2], Purusotam Basnet[3,5], Ganesh Acharya[5,6], Paramasivam Senthilkumaran[7], Dalip Singh Mehta[1, 7] and Balpreet Singh Ahluwalia[2,6]

[1] *Bio-photonics Laboratory, Department of Physics, Indian Institute of Technology Delhi, Hauz-Khas, New Delhi- 110016, India.*
[2] *Department of Physics and Technology, UiT The Arctic University of Norway, Tromsø, (Norway).*
[3] *Women´s Health and Perinatology Research Group, Department of Clinical Medicine, UiT The Arctic University of Norway, Tromsø, Norway.*
[4] *Department of Computer Science, UiT The Arctic University of Norway (Norway).*
[5] *Department of Obstetrics and Gynaecology, University Hospital of North Norway, Tromsø, (Norway).*
[6] *Department of Clinical Science, Intervention and Technology Karolinska Institutet, Stockholm, Sweden.*
[7] *Department of Physics, Indian Institute of Technology Delhi, Hauz-Khas, New Delhi- 110016, India.*
Email: *ankitbutola321@gmail.com*
**\*Corresponding author:** *balpreet.singh.ahluwalia@uit.no*
#: These authors contributed equally in this work.

**Abstract:**
Sperm cell motility and morphology observed under the bright field microscopy are the only criteria for selecting particular sperm cell during Intracytoplasmic Sperm Injection (ICSI) procedure of Assisted Reproductive Technology (ART). Several factors such as, oxidative stress, cryopreservation, heat, smoking and alcohol consumption, are negatively associated with the quality of sperm cell and fertilization potential due to the changing of sub-cellular structures and functions which are overlooked. A bright field imaging contrast is insufficient to distinguish tiniest morphological cell features that might influence the fertilizing ability of sperm cell. We developed a partially spatially coherent digital holographic microscope (PSC-DHM) for quantitative phase imaging (QPI) in order to distinguish normal sperm cells from sperm cells under different stress conditions such as cryopreservation, exposure to hydrogen peroxide and ethanol without any labeling. Phase maps of 10,163 sperm cells (2,400 control cells, 2,750 spermatozoa after cryopreservation, 2,515 and 2,498 cells under hydrogen peroxide and ethanol respectively) are reconstructed using the data acquired from PSC-DHM system. Total of seven feedforward deep neural networks (DNN) were employed for the classification of the phase maps for normal and stress affected sperm cells. When validated against the test dataset, the DNN provided an average sensitivity, specificity and accuracy of 84.88%, 95.03% and 85%, respectively. The current approach DNN and QPI techniques of quantitative information can be applied for further improving ICSI procedure and the diagnostic efficiency for the classification of semen quality in regards to their fertilization potential and other biomedical applications in general.

**Introduction:**
Semen quality and male fertility potential have been continuously declining all over the world [1-4]. At the same time, biomedical and technical advances have made it possible to treat male infertility using assisted reproductive technology (ART) including intracytoplasmic sperm injection (ICSI). Evaluation of semen quality and ICSI procedure are the important steps for the successful outcome of ART. Generally, semen parameters evaluation and ICSI procedure are guided by the bright field microscopy and experience of laboratory person. Recently, computer-assisted sperm analysis (CASA) a digital microscopic technique made it possible as a machine-based analysis for semen parameters such as sperm cell concentration, sperm cell motility, kinematics, and morphology. For instance, CASA systems acquire successive images of the cells and use special software to track the motion of heads of each spermatozoon [5, 6]. However, it fails to provide any supportive information regarding to subcellular changes within the sperm cells which could be useful to ICSI procedure. Another powerful technique is a label free holographic imaging, for 3D reconstruction of freely moving sperm cells [7]. Various optical as well as spectroscopic techniques have been proposed so far to determine motility of the sperm cells [8-11]. For example, fluorescence imaging and laser scanning confocal microscopy to investigate the mitochondrial functionality of sperm cells [12] since the motility of the cells is partially depends on the mitochondrial function [13] and get affected by cryopreservation of the cells. Semen quality is also analyzed based on Raman micro spectroscopy which can provide the spectral features of human sperm cells [14]. Oxidative stress is also known to affect the integrity of sperm genome, result in lipid peroxidation and decrease in sperm motility, which was quantified recently using partially spatially coherent digital holographic microscopy and machine learning [15]. Additionally, smoking and alcohol consumption are negatively associated with sperm concentration and percentage of motile sperms when compared with the persons without these habits [16]. All these factors can affect the morphology, physiology and subcellular structures of the sperm cells and these changes can be investigated by extracting the quantitative information of the cells

For label-free sperm imaging, quantitative phase imaging (QPI) is an attractive non-invasive technique to extract the quantitative information of the samples [17-21]. QPI can measure the combined information of refractive index and thickness of the specimens with a nanometric sensitivity, which can be utilized detecting any deviations from normality [22-24]. QPI has several biological applications, such as 3D imaging of human red blood cells (RBC) [25], bovine embryo [26], bovine spermatozoa [27], tissue imaging [28], and others. Although, QPI is a potentially useful technique as it provides phase map in terms of height and refractive index of the samples, it has not yet been amenable to interpretation and classification by human expert due to the lack of specificity [29]. Therefore, merging QPI with artificial intelligence (AI) is a promising route to provide virtual image classification of the QPI data [29]. Recently, AI techniques have been implemented to differentiate healthy and non-healthy sperm cells. For example, support vector machine (SVM) technique uses morphological and textures features of the cells to classify them into diagnostically relevant classes [15, 30, 31]. Although it provides good classification accuracy, manual selection of relevant texture features out of thousands of features is the main challenge with conventional machine learning technique. In contrast, deep learning-based classification does not require to extract any features and can automatically generate abstract convolutional features from the training dataset. Deep learning is rapidly growing as an automated technique in biomedical imaging for example disease classification [32-34], image segmentation [35], resolution enhancement [36], digital staining [18], noise reductions [37], among others[38, 39].

We demonstrate the use of QPI technique assisted with deep learning for the classification of sperm cells under different stressed conditions. A total of four different classes were considered in this study, which included healthy, externally induced oxidative stressed, cryopreserved, and externally induced alcohol affected sperm cells. The four classes of sperm cells were also studied using conventional techniques to quantify and compare the progressive and the non-progressive motility of spermatozoa. Figure 1 shows the schematic representation of a partially spatially coherent digital holographic microscope (PSC-DHM) developed to acquire the interferometric images of the sperm cells as can be seen in Fig. 1(b). Figure 1(c) and (d) shows the quantitative phase map of sperm cells. The PSC-DHM system offers single shot phase reconstruction of the cells especially the thinnest i.e. tail part of the sperms by utilizing partial spatial coherent properties of light source. QPI system commonly uses direct laser light (high spatially and temporally coherent) or white light (spatially and temporally incoherent). Direct laser suffers with low spatial phase sensitivity and hence accurate phase estimation of thin sample is usually difficult. On the other hand, white light provide high spatial phase sensitivity but due to low temporal coherence it requires phase shifting technique to utilize whole field of view of the camera. In contrast, PSC-DHM system offers single shot phase extraction of the sample due to high temporal coherence and, high spatial phase sensitivity because of its spatial incoherent nature. The details of the PSC-DHM system is provided in the Methods section. The spatial phase sensitivity of the system developed is around ±20 mrad which is utilized to reconstruct the phase information of the cells including the tail part. The thickness of tail of the sperm cells are typically 100 nm and thus more challenging to image unless high spatial phase sensitivity is achieved. This is particularly useful as the tail plays an important role in progressive motility of the cells and it may change under different stressed conditions.

In this study, total 10,163 interferometric images (2,400 normal, 2,750 cryopreserved, 2,515 oxidative stressed and 2,498 alcohol affected) of the sperm cells were acquired from the PSC-DHM system. The phase maps of these cells were used as inputs to the DNN which was trained by 70% of the data and validated against the testing data to check the classification accuracy of the networks. The current QPI+DNN approach has several advantages over the conventional techniques for sperm classification. First, QPI allows to extract the quantitative information that is directly related to the morphology of the sperm cells and is otherwise not possible by bright field microscopy. Second, the DNN provides automated classification of phase map of the cells. We believe QPI coupled with DNN would find usage in IVF clinics in diagnosis, and for selection of healthy cells.

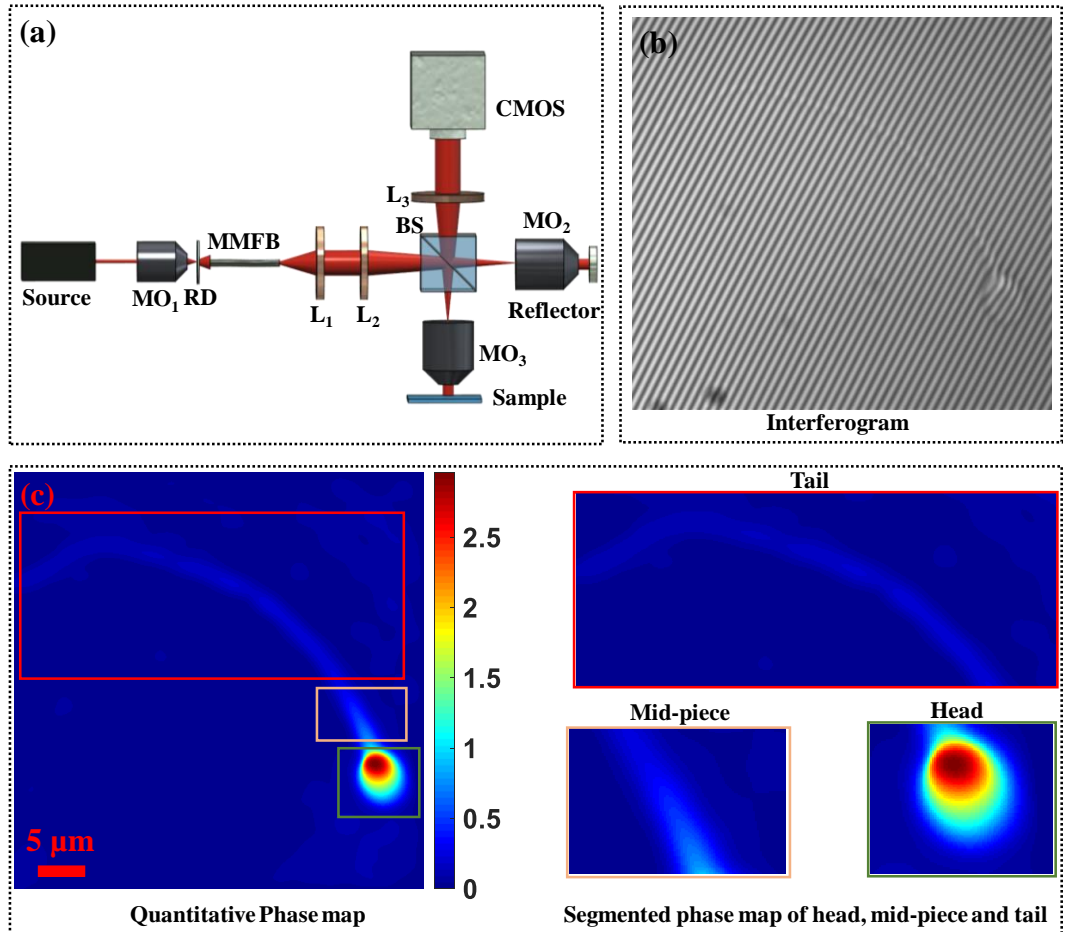

**Figure 1**: Schematic diagram of the partially spatially coherent digital holographic microscope (PSC-DHM) system (a) and the interferometric image of the sperm cell acquired from the PSC-DHM (b). (RD- rotating diffuser, L- lens, BS- Beam splitter, MO-microscope objective, MMFB- multiple multi-mode fiber bundle). Reconstructed phase map (c) and the zoomed view of head, neck and tail part of the sperm cell. Color bar represents the phase map in radian.

**Results and discussion:**

**Motility test of spermatozoa exposed to ethanol, hydrogen peroxide, and cryopreservation:**

The effect of ethanol, hydrogen peroxide and cryopreservation on progressive and non-progressive motility of spermatozoa was investigated and the results are shown in Fig. 2. For externally induced stress experiments a fixed concentration, i.e. ethanol (2%) or hydrogen peroxide (200 μM $H_2O_2$) were used to simulate stress condition and to observe functional and morphological changes of the sperm cells. Sperm cells were exposed to ethanol or hydrogen peroxide for 1 hour at 37 °C. For the control group, same amount of medium as in the test group were added.

Figure 2 shows the box plot of the sperm cells after different stressed conditions. After incubation for 1 h, ethanol produced a significant decrease in progressive motility of sperm cells as compared with control (Table 1): 18.7 ± 13.8% vs control 73.9 ± 19.5%, at the same time after incubation with ethanol the percentage of non-progressive motile cells increased: 33.1±11.94% vs control 14.6±13.8% (ANOVA, paired t-test, $p < 0.05$). Similarly, $H_2O_2$ had significantly decreased the progressive motility of sperm cells compared with controls (Fig. 2, Table 1): 2.4±4.04% vs control 73.9 ± 19.5%. Incubation with $H_2O_2$, the non-progressive motility was significantly decreased as compared with control: 77.7 ± 16.2% vs control 14.6±13.8% (ANOVA, paired t-test, $p < 0.05$). Cryopreservation resulted in a significant decrease in progressive motility (Fig. 2, Table 1). Spermatozoa with progressive motility had a mean value before cryopreservation of 73.9%, while mean value after thawing was

17.3% (p < 0.01). The non-progressive motility decreased after thawing, but not significantly: 27.1±19.5% vs 14.6±13.8% (p > 0.01).

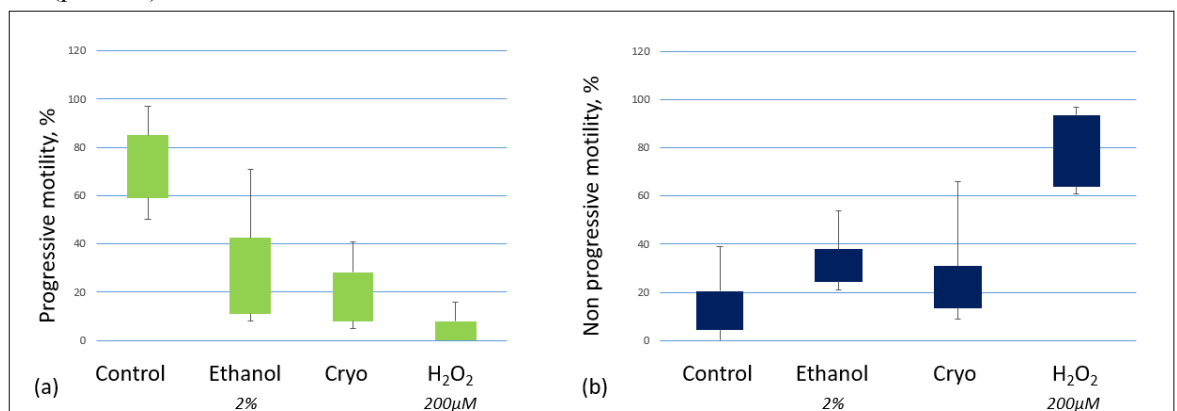

**Figure 2.** Progressive (a) and non-progressive (b) motility changes of sperm cells after incubation (1h/37°C) with ethanol, hydrogen peroxide ($H_2O_2$) and after cryopreservation as compared with control (mean ± SD, p<0.05 vs. control), (n = 7; seven ejaculates from different donors).

**Table 1.** Effect of cryopreservation, ethanol and hydrogen peroxide incubation on human sperm cells motility. Analysis of the differences among group means using Paired Two Sample t-Test for Means (alpha 0,05). Values are shown as a mean ± standard deviation (SD). "**a, b and c**" – Mean differences between ethanol and control group, "**b**" – between $H_2O_2$ and control group, "**c**" – between cryosample and control.

| Variable | Ethanol, 2% Mean±SD | $H_2O_2$, 200µM Mean±SD | Cryopreservation Mean±SD | Control Mean±SD | P-Value |
|---|---|---|---|---|---|
| Progressive motility (PR, %) | 18.7 ± 13.8 | 2.4±4.04 | 17.3±11.9 | 73.9 ± 19.5 | $X^a$ 0.00009 $Y^b$ 0.00005 $Z^c$ 0.0001 |
| Non-progressive motility (NP, %) | 33.1±11.94 | 77.7 ± 16.2 | 27.1±19.5 | 14.6±13.8 | $X^a$ 0.01 $Y^b$ 0.0002 $Z^c$ 0.2 |

A possible mechanism of decrease in sperm motility after treatment with ethanol is distortion of cell membrane caused by altering of membrane protein structure [40, 41]. On the other hand, the principle mechanism of the effect of hydrogen peroxide on sperm motility is peroxidation of unsaturated fatty acids, which is a part of membrane lipids. As a result of peroxidation, the membrane loses flexibility and plasticity which determines disrupted tail motion [42]. The functional changes of sperm cells after cryopreservation might be influenced by oxidative stress, cryo-protector used (type and concentration), methods of cryopreservation and thawing itself, and is also dependent on original semen characteristics, such as concentration and motility [42]. In accordance with the concept of «partial survival» current procedures used for sperm freezing and thawing lead to decrease of more than 50% in motility and survival rate [43], ultrastructure and cell morphology changes [44-46], mitochondrial activity reduction [47], and damages of sperm chromatin [42].

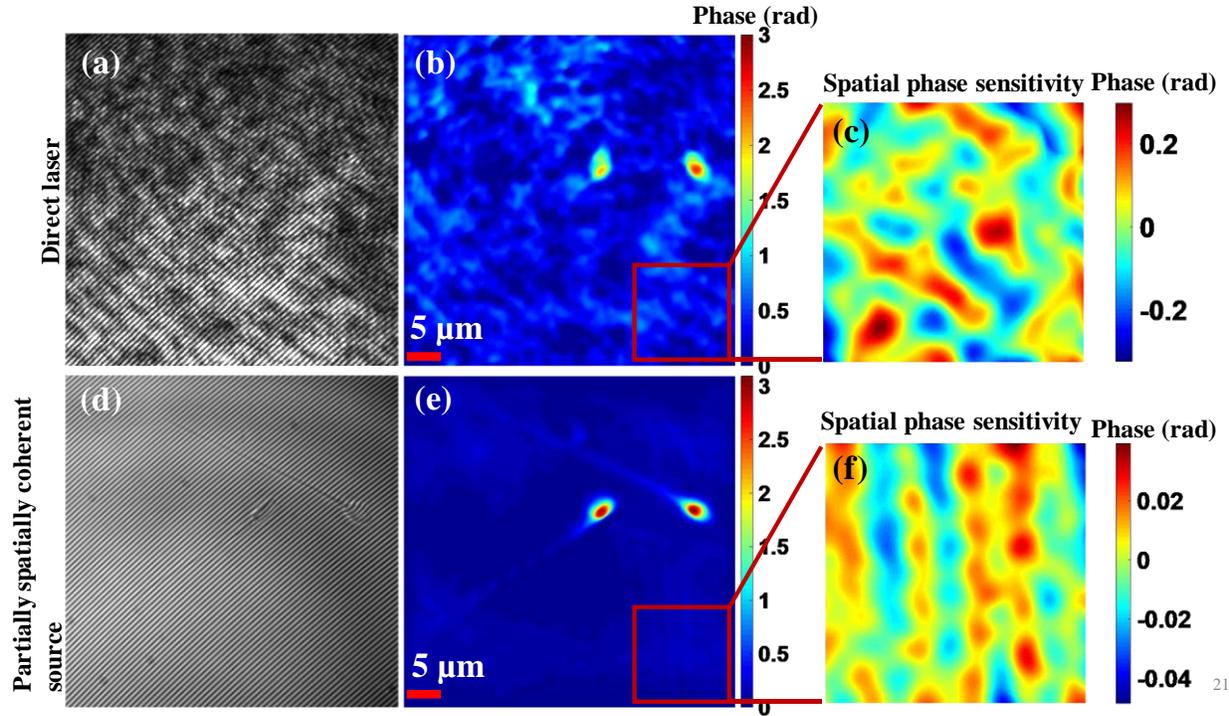

**Figure 3:** Comparison between direct laser and partially spatially coherent light source for the quantitative phase imaging of sperm cells. Figure (a) and (b) represents the interferogram and reconstructed phase map of the sperm cells in case of direct laser. Figure (d) and (e) shows the interferogram and phase map of same location using partially spatially coherent light source where mid-piece and tail part are clearly observed. Spatial phase sensitivity represents the phase noise of the system.

Figure 3 depicts the phase reconstruction using direct laser and partially spatially coherent light source. Speckles and spurious fringe pattern degrade the quality of interferogram (Fig. 3(a)) and thus poor spatial phase sensitivity (±200 mrad) in case of direct laser. Spatial phase sensitivity relates with phase measurement accuracy of the system i.e. larger the value, poorer will be the accuracy. Spatial phase sensitivity of partially spatially coherent light source (Fig. 3 (d-f)) can be seen 10 times higher than the direct laser and hence phase reconstruction of thinnest part of the sperm i.e. tail is clearly seen. Figure 4 shows a schematic of our framework for the selection of normal sperm cells. The interferometric image of a sperm cell acquired from PSC-DHM as shown in Fig 1 (b), is processed using standard Fourier transform algorithm [48] to extract the wrapped phase map of the cell. Phase unwrapping is done using Goldstein algorithm [49]. The details of the Fourier transform and steps for reconstruction of phase map are described in the Methods section. Recently, classification of the phase map of sperm cells is performed using simple machine learning techniques [15] where texture and morphological features are extracted from the phase image. These features are fed into machine learning models to separate the corresponding phase images into their relevant classes [30]. In contrast, in this work we develop an end-to-end deep learning approach for the classification of healthy sperm cells, which does not require extraction of any features from the phase image. The DNN takes a phase image as input and provide a diagnostically relevant class label as an output i.e. normal cells, $H_2O_2$ stressed cells, ethanol stressed cells, and cryopreserved cells. DNN architecture consists of difference combination of convolution layer, rectifier linear unit layer, maxpooling layer, fully connected layer and finally softmax layer. Details of these layers can be found elsewhere [50]. A total of seven deep neural networks, namely AlexNet, GoogLeNet, Inception-ResNet-V2, VGG-16, VGG-19, ResNet-50 and ResNet-101 have been investigated. These networks are trained by total 70% of the phase images and 30% for testing the accuracy of each model. Accuracy of these network are shown in terms of confusion matrices. Confusion matrix shows the number of correct and incorrect prediction of the network.

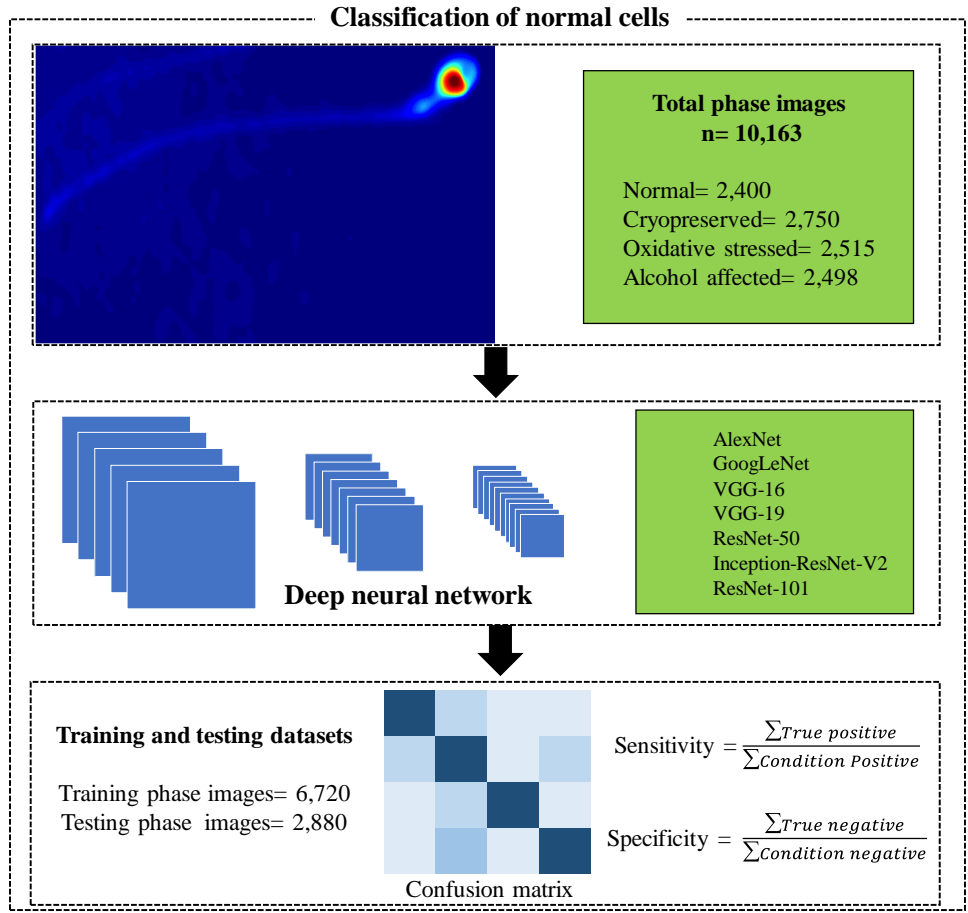

**Figure 4:** Workflow diagram showing the important steps for the classification of quantitative phase map of sperm cells. Phase map of the images is reconstructed by the interferogram captured using PSC-DHM system. Classification of the phase images is done by deep neural networks (DNN) and the output of the network shows in accuracy and sensitivity of the testing datasets in terms of confusion matrix.

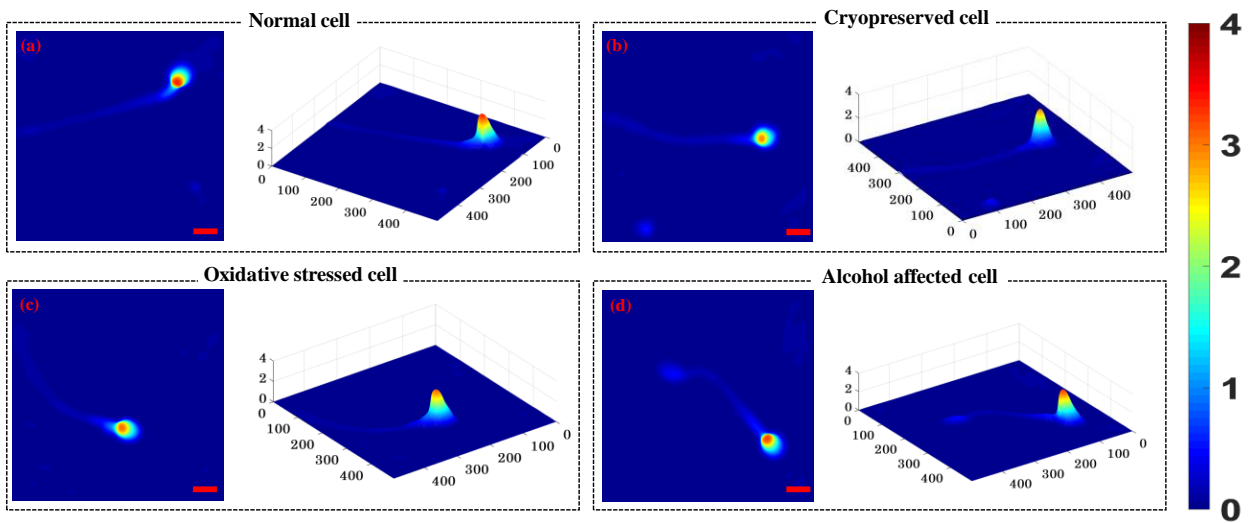

**Figure 5:** Quantitative phase map of human sperm cells, reconstructed from the interferogram captured by PSC-DHM system: (a) normal cell, (b) after cryopreservation, (c) oxidative stressed cell and (d) alcohol affected cell. Color bar represents the phase map in radian. Scale bar: 5 μm.

Figure 5 represents the reconstructed quantitative phase maps of human sperm cells under different stressed conditions. All interferometric images were acquired using 60X 1.2NA (UPLSAPO 60XW, Olympus) microscopic objective lens. Figure 5 (a-d) are the quantitative phase maps of normal, cryopreserved, externally introduced ethanol and externally introduced $H_2O_2$ sperm cells respectively. The scale bar corresponds to 5 μm distance in both x and y directions and the color bar shows the phase value in radians. A total 12,332 phase images (2,906 healthy, 2,981 cryopreserved, 3,222 ethanol and 3,223 oxidative stressed) are reconstructed from the interferometric images acquired from the proposed setup. Among the 12,332 images, only those phase images are retained that follow the following criteria: correct phase unwrapping, background subtraction and only one cells lies throughout the field of view. The first criterion allows for filtering of phase images that are reconstructed correctly and the later criterion is to promote accurate classification of the sperm cells using DNN. Selection of images with single cells is done automatically by converting phase image into a binary image and setting certain threshold value of white pixels. Threshold value was selected empirically by determining the threshold that gives reasonable segmentation of the cells in the binary image of a candidate image. For the empirical threshold determination, a small set of images with only one cell each were hand-picked as the candidate images. The process of thresholding generally filtered away most images with multiple cells. Nonetheless, all the retained images were checked manually to assess if both the above-mentioned criteria were satisfied, so that the retained images are indeed suitable for further use. A total of 10,163 phase images of sperm cells (2,400 healthy, 2,750 cryopreserved, 2,515 oxidative stressed and 2,498 alcohol affected) are thus retained.

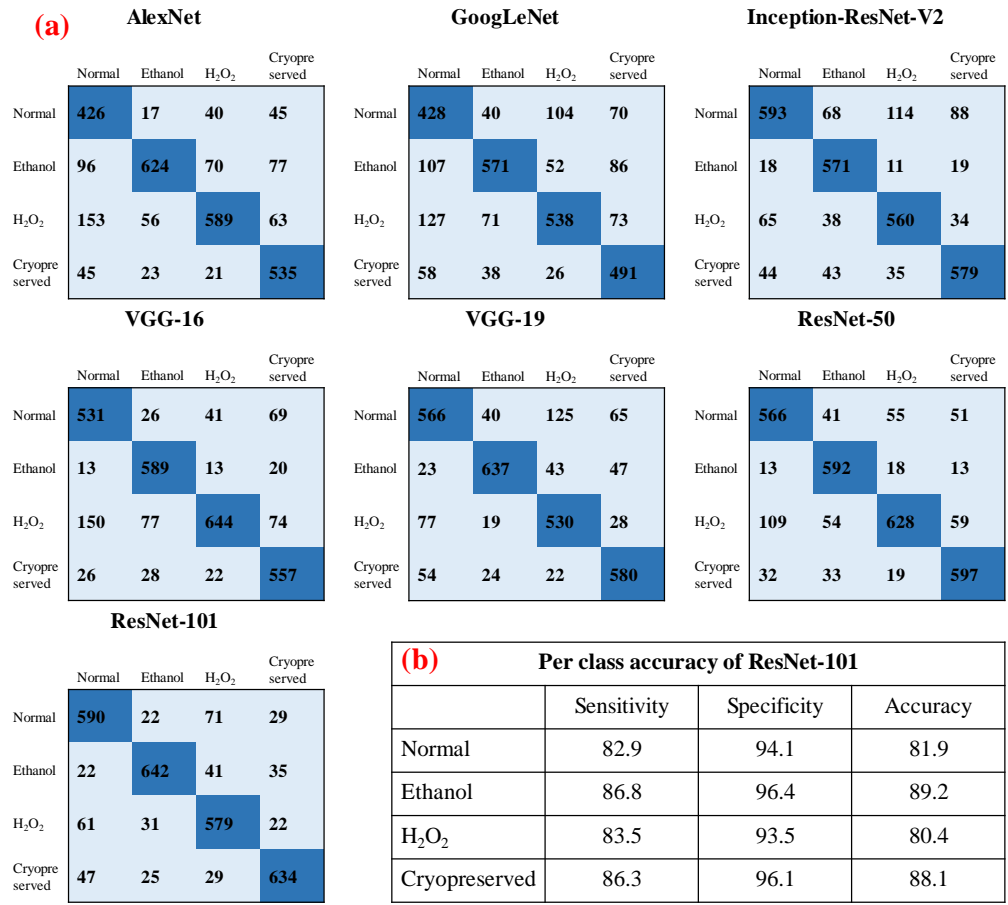

**Figure 6:** Performance of the deep neural networks (DNN) on the testing datasets of the phase images of sperm cells. (a) Confusion matrices of different DNN with number of phase images for classification of healthy and non-healthy phase map of sperm cells. Diagonal elements show number of correct predictions and the off-diagonally elements are the wrong classified observations. (b) Per-class sensitivity, specificity and accuracy of ResNet-101.

From Fig. 5, the quantitative phase map of healthy sperm's head is found to be maximum as compared to the cells under different stressed conditions. The scale bar represents the phase map (thickness+ refractive index) in radian. Deep red indicates the maximum phase and the deep blue corresponds to zero phase. Change in phase value might indicate the change in morphology of the head of the cells under different stressed condition. However, no general trend in the maximum phase map of the cells is observed which can be explained from the progressive/non-progressive motility and number of mobile (Fig. 2) cells. Although, the number of mobile sperms get decreased as compared to the normal class, there are still some cells which are mobile. Thus, they may sustain their morphology close to normal cell and hence no general trend of maximum phase map is observed between these classes.

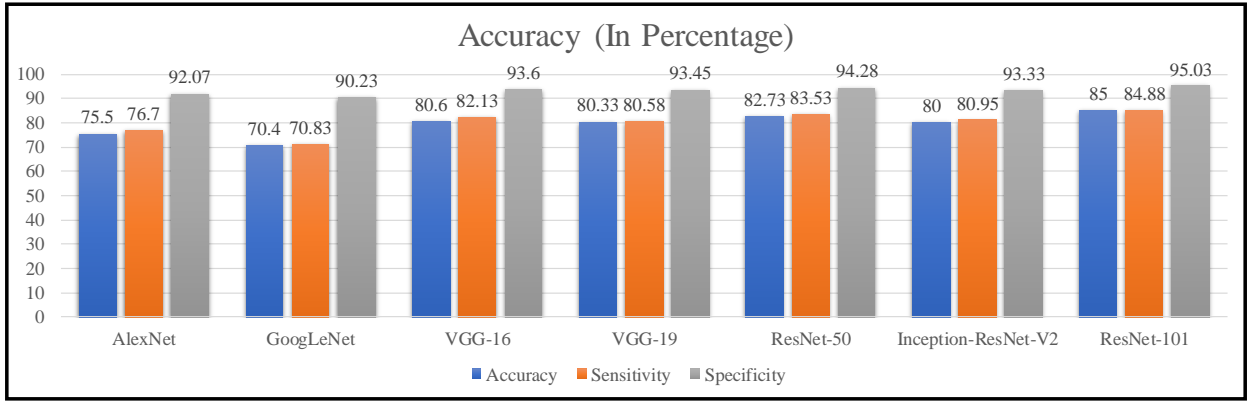

**Figure 7:** Sensitivity, specificity and classification accuracy of different deep neural network. The green bar showing the accuracy of each architectures and the ResNet-101 is providing the best accuracy (85%) on the testing datasets.

The performances of the different DNN architectures used in this study are reported in Fig. 6. Total 70% of the data is used for the training purpose while 30% is used to test the accuracy of each model. It is important to note that sufficient training is necessary to train the network and to achieve best accuracy while validating against the test data. Figure 6 (a) represents the confusion matrices of the DNN on the testing datasets of the phase image of sperm cells. Confusion matrices shows the performance of the network against the testing dataset. Rows and column of the matrices indicate the predicted class and the ground truth respectively. Diagonal elements of the matrix show the correct predictions of the data while the off-diagonal elements are the wrong classified data.

Consider the confusion matrix of AlexNet in Fig. 6 (a). Total 528 phase images of the normal phase image of sperm cells are tested by the network. Out of 528, 426 phase images are predicted correctly and 17, 40 and 45 are the wrong predictions into the classes corresponding to ethanol, $H_2O_2$ and cryopreservation stressed cells respectively. Similarly, 624, 589 and 535 are the correct prediction by the AlexNet for Ethanol, $H_2O_2$ and cryopreserved cells, respectively. Performance of the GoogLeNet, Inception-ResNet-V2, VGG-16, VGG-19, ResNet-50 and ResNet-101 can be also seen in the confusion matrix. Note that, the total test images are 30% of the total image (n=9600) i.e. 2880 which is summation of all the elements presents in the matrix. Figure 7 depicts the average sensitivity, specificity and classification accuracy of all deep learning architectures. Out of all DNN architectures, ResNet-101 provides the best sensitivity, specificity and accuracy of 84.88%, 95.03%, and 85% respectively.

**Conclusion:**

Our current QPI+DNN framework allows for the automated classification between normal and abnormal sperm cells. QPI is a promising technique which provides phase map as a function of refractive index and thickness of the sample and hence can have an edge over the conventional intensity-based identification of healthy cells. PSC-DHM system can be used to extract the quantitative phase map of the sperm cells enabling single shot phase reconstruction with high spatial phase sensitivity (±20 mrad). High spatial phase sensitivity is utilized to acquire the phase map of the entire sperm cells, i.e. head, neck and tail of the sperm cells, which is difficult to image using direct laser based DHM [15]. Sperm cells after cryopreservation, oxidative stress and exposure to ethanol can be imaged by the proposed setup. Previous studies show that oxidative stress initiates concentration dependent increase of DNA fragmentation because of DNA strand breaks [40]. Also, ethanol distorted the cell membrane resulting from the alteration of membrane protein structure [4]. Therefore label-free, non-invasive methods such as PSC-DHM are highly desirable to detect these changes and hence for the selection of good quality sperm for ICSI procedure that can be used to improve the success of ART however it need further clinical trails.

Quantitative analysis of sperm cells provides an opportunity to identify healthy sperm cells using deep learning approaches. Deep learning can be potentially a powerful technique for automated classification of sperm cells into normal and abnormal. Our results demonstrate that a variety of DNN architectures for classification provide good accuracy of classification into 4 classes. ResNet-101 provided the best accuracy, i.e. 85%, for classification into healthy, oxidative stressed, cryopreserved and ethanol affected sperm cells. Moreover, the use of seven different network allows us to understand the capabilities of each networks and to apply best deep learning architecture for the identification of healthy cells. We applied our automated classification model for studying clinically relevant problems of semen quality in different patients attending the *in-vitro* fertilization clinic of University Hospital of North Norway, Norway. Fully automated classification of the sperm cells could be an intermediate tool for the expert that can be utilized for the selection of healthy sperm cells as per WHO criteria [51]. QPI + DNN framework for healthy sperm identification could be potentially used for real time selection of healthy living sperm cells that can be used for improving the success of fertilization during ART procedures.

**Methods:**
**Experimental setup:**

PSC-DHM is developed for quantitative phase imaging (QPI) of the sperm cells. Schematic diagram of PSC-DHM setup is shown in Fig. 1. To reduce the spatial coherence of the direct laser (He-Ne laser), it first focused by using microscopic objective lens ($MO_1$) and rotating diffuser is placed at the focus plane of the $MO_1$. Rotating diffuser scattered the light into multiple directions which captured by multi-multimode fiber bundle (MMFB). Output of MMFB consist high temporally and low spatial coherence properties and thus act as an extended light source. Details of such type of light source can be found elsewhere [52]. The extended light source coupled at an input port of the Linnik type inteferometer. The light beam is first collimated and then focused by using a combination of lens $L_1$ (f= 50 mm) and $L_2$ (f= 150 mm) respectively. A beam splitter is placed to divide the focused beam into reference and sample. The light beam is focused into the back focal plane of $MO_3$ (UPLSAPO 60XW, Olympus) and hence the output beam is nearly collimated to extract the accurate phase information of the sample. The light beam reflected from the sample and reference mirror, interfere at the beam spliiter plane which consist the coded phase information of the sample. The interferogram is finally projected into the camera sensor (Hamamatsu ORCA-Flash4.0 LT, C11440-42U) by using a tube lens $L_4$ (f= 150 mm). The 2D intensity variation of an interferogram can be expressed as:

$$I(x,y) = a(x,y) + b(x,y)\cos[2\pi i(f_x x + f_y y) + \phi(x,y)] \quad (1)$$

where $f_x$ and $f_y$ are the spatial carrier frequencies of the interferogram. Background (DC) and the modulation terms are defined by $a(x,y)$ and $b(x,y)$, respectively. Phase $\phi(x,y)$ contains information of the specimen. By applying Fourier transform, the phase map of the specimen can be measured by the following expression:

$$\phi(x,y) = \tan^{-1}\left[\frac{Im(c(x,y))}{Re(c(x,y))}\right] \quad (2)$$

where

$$c(x,y) = b(x,y)\exp(i\phi(x,y)) \quad (3)$$

where Im and Re are the imaginary and real part of the complex signal. As the wrapped phase map lies between $-\pi$ to $+\pi$, the unwrapping is done by standard Goldstein phase unwrapping algorithm [49].

**Semen preparation.** The Regional Committee for Medical and Health Research Ethics of Norway (REK_nord) has approved the study. Ethical guideline were followed. At the IVF Clinic, Department of Obstretics and Gynecology, University Hospital North Norway, Tromsø, 7 semen samples were collected from patients who were attended the IVF clinic for the service of ART. All patients were informed and consent was obtained. The semen sample was collected according to the criteria established by the WHO [51] after 3-5 days of abistance. After liquefaction, sperm counts were evaluated using the Neubauer-improved counting chamber. All ejaculates used in the experiments meet as the good quality semen sample as per requirements of WHO 2010 (Table 2). To eliminate seminal plasma and isolate cells with good quality sperm cells, one milliliter of semen was carefully placed on the each 1.5 ml of 90% and 45% gradient layers (Vitrolife, Sweden) and centrifugated at 500 g for 20 min. The resultant pellet was washed twice with human Quinn's sperm washing medium (Origio, Denmark) at 300 g for 10 min. The supernatant was discarded, and the concentration of the cells from the pellet was adjusted to $1x10^6$ sperm per mL with Quinn's Advantage fertilization medium (Origio, Denmark) supplemented with 5 mg/ml Human Serum Albumin (Sigma).

To perform experiments, 96-well cell culture plates (Corning) were filled with purified sperm in a concentration of $2x10^4$ cells per mL with 200 μM $H_2O_2$ (for oxidative stressed samples) or 2% ethanol (for alcohol affected samples), the reference chamber was filled with purified semen only. The samples were incubated for 1 hour at 37°C, 5% $CO_2$. After incubation evaluation of cell motility was graded according to WHO 2010 criteria as a progressive (PR) and non-progressive motility (NP). Sperm counting was performed using Neubauer-improved counting chamber and examined under the inverted phase contrast microscope at 40x magnification. Cryopreservation and thawing of purified semen were performed in accordance with Sperm Freeze medium protocol (Origio, Denmark). Motility of post-thaw spermatozoa was evaluated using Neubauer-improved counting chamber. For the purpose of quantitative analysis by PSC-DHM the cells of each sample were transferred in a PDMS chamber on reflecting silicon. Sperm cells were immobilized by fixation with 4% PFA for 30 min at RT and washed in PBS (Phosphate-Buffer Saline, Sigma) for 5 min. Finally, fixed and attached at the surface of PDMS chamber sperm cells were mounted in PBS and covered by the cover glass of 170 micron thickness.

| Parameter | Mean ± SD |
|---|---|
| Age (years) | 34.7±4.83 |
| Semen volume (ml) | 3.12±1.46 |
| Semen concentration ($x10^6$/ml) | 51.6±22.84 |
| Total sperm count ($x10^6$) | 166.79±142.12 |
| Progressive motility (%) | 59±11.97 |

**Table 2.** Age and semen quality measured before the purification by gradient method (n=7, number of donors). Values are shown as a mean ± standard deviation (SD).

**Data analysis:**
Extracting phase information from the interferogram and deep learning is implemented in Matlab® 2019a on a 64-bit Windows OS, Intel Xeon CPU E5-1650 v4 @ 3.6 GHz with 64 GB RAM and Nvidia 2080 Ti GPU. Transfer learning is performed by using pretrained DNN and retraining them for our desired classes. Classification results are obtained by randomly assigning 70% of the images in the dataset for training and the remaining 30% for testing. In each training iteration, the initial learning rate is set as $10^{-4}$ and stochastic gradient descent with momentum (SGDM) is used for training. Maximum number of epochs in the learning process is set as 20.

**Acknowledgement:** B.S.A acknowledges the funding from the Norwegian Centre for International Cooperation in Education, SIU-Norway (Project number INCP- 2014/10024) and from the Research Council of Norway, (Nano2021–288565). We are grateful to the patients who agreed to provide semen samples and bioengineers (Sissle A. Hansen, Inger K. Olaussen) who recruited the patients to collect the samples.

**Author contributions:** BSA, DSM, PB, GA and PS discussed the idea. The samples were provided by PB, who also sought the permission and ethical clearances for this study. PB, DP and GA provided biological insights for this study. Samples were prepared by DP. PSC-DHM system is developed by AB, AA and JCT. AH designed and optimize the rotating diffuser and provide the sample substrate to perform the experiment. Experimental data are acquired by AB. The support for DNN architectures, training, and performance evaluation was provided by DKP in close collaboration with AB. AB and DP prepared first draft of the manuscript and all author contributed towards writing of the manuscript. The work is supervised by PB and BSA.

**Competing interests:** Authors declare no competing interest.